\documentstyle[12pt,epsf]{article}

\begin{document}

\begin{flushright}
SLAC--PUB--7115\\
March 1996
\end{flushright}

\bigskip\bigskip
\begin{center}
{\bf\large
DISCRETE PHYSICS AND THE DIRAC EQUATION\footnote{\baselineskip=12pt
Work partially supported by Department of Energy contract DE--AC03--76SF00515
and  by the National Science Foundation under NSF
Grant Number DMS-9295277.}} 

\bigskip

Louis H. Kauffman\\
Department of Mathematics, Statistics and Computer Science\\
University of Illinois at Chicago\\
851 South Morgan Street, Chicago IL 60607-7045\\[2ex]
and\\[2ex]
H. Pierre Noyes\\
Stanford Linear Accelerator Center\\
Stanford University, Stanford, CA 94309\\
\end{center}
\vfill
\begin{abstract}
We rewrite the 1+1 Dirac equation in light cone coordinates in two
significant forms, and solve them exactly using the classical
calculus of finite differences. The complex form yields ``Feynman's
Checkerboard''---a weighted sum over lattice paths. The rational,
real form can also be interpreted in terms of bit-strings.
\end{abstract} 
\vfill
\begin{center}
Submitted to Physics Letters {\bf A}.
\end{center}
\vfill
{\baselineskip 12pt
\noindent

 \par}
\vfill
\newpage

\section{Introduction}

In this paper we give explicit solutions to the Dirac equation for
1+1 space-time. These solutions are valid for discrete physics
\cite{Discrete} using the calculus of finite differences, and they
have as limiting values solutions to the Dirac equation using
infinitesimal calculus. We find that the discrete solutions can be
directly interpreted in terms of sums over lattice paths in discrete
space-time. We document the relationship of this lattice-path with
the checkerboard model of Richard Feynman \cite{quantum}. Here we
see how his model leads directly to an {\it exact} solution to the
Dirac equation in discrete physics and thence to an exact continuum
solution by taking a limit. This simplifies previous approaches to
the Feynman checkerboard \cite{Jacobson,Karmanov}.

We also interpret these solutions in terms of choice sequences
(bit-strings) and we show how the elementary combinatorics of
$i=\sqrt{-1}$ as an operator on ordered pairs ($i[a,b] =[-b,a]$)
informs the discrete physics. In this way we see how solutions to
the Dirac equation can be built using only bit-strings, and no
complex numbers. Nevertheless the patterns of composition of $i$
inform the inevitable structure of negative case counting
\cite{Etter,EtterFinal} needed to build these solutions.

The paper is organized as follows. Section 2 reviews the Dirac
equation and expresses two versions (denoted RI, RII) in light cone
coordinates. The two versions depend upon two distinct
representations of the Dirac algebra. Section 3 reviews basic facts
about the discrete calculus and gives the promised solutions to the
Dirac equation. Section 4 interprets these solutions in terms of
lattice paths, Feynman checkerboard and bit-strings. Section 5
discusses the meaning of these results in the light of the
relationship between continuum and discrete physics.
\vfill
\newpage

\section{The 1+1 Dirac Equation in Light Cone Coordinates}

We begin by recalling the usual form of the Dirac equation for one
dimension of space and one dimension of time. This is
\begin{equation}
i\hbar\ \frac{\partial \psi}{\partial t} = E\psi
\end{equation}
where the energy operator $E$ satisfies the dictates of special
relativity and obeys the equation
\begin{equation} 
E= c\sqrt{p^2 + m^2c^2}
\end{equation}
where $m$ is the mass, $c$ the speed of light and $p$ the momentum.
Dirac linearized this equation by setting $E=c\alpha p + \beta mc^2$
where $\alpha$ and $\beta$ are elements of an associative algebra
(commuting with $p$, $c$, $m$). It then follows that
\begin{equation}
c^2(p^2+m^2c^2) =(c\alpha p + \beta mc^2)^2=c^2p^2\alpha^2 +
m^2c^4\beta^2 + c^3pm(\alpha \beta + \beta\alpha) \ .
\end{equation}
Thus whenever $\alpha^2=\beta^2=1$ and $\alpha\beta+\beta\alpha=0$,
these conditions will be satisfied. Thus we have Dirac's equation in
the form $ i\hbar\ \frac{\partial \psi}{\partial t} = (c\alpha p
+\beta mc^2)\psi$. For our purposes it is most convenient to work in
units where $c=1$ and $\hbar/m=1$. Then $i\ \frac{\partial\psi}
{\partial t}=(\alpha p/\hbar +\beta )\psi$ and we can take
$p=\frac{\hbar}{i}\, \frac{\partial}{\partial x}$ so that the
equation is
\begin{equation}
i\, \frac{\partial \psi}{\partial t} = \left(-\alpha i\,\frac{\partial}
{\partial x} +\beta \right)\psi \ . 
\end{equation}

We shall be interested in $2\times 2$ matrix representations of the
Dirac algebra $\alpha^2=\beta^2=1$, $\alpha\beta+\beta\alpha=0$. In
fact we shall study two specific representations of the algebra. We
shall call these representations RI and RII. They are specified by
the equations below
\begin{eqnarray}
 RI: \ \ \ \alpha &=& 
           \left(\begin{array}{cc}  -1 & 0\\  0 & 1\end{array}\right) ,\ \
   \beta = \left(\begin{array}{cc}  0  & -i\\ i & 0\end{array}\right)\\[2ex]
 RII: \ \ \ \alpha &=&
           \left(\begin{array}{cc} -1 & 0\\  0 &1\end{array}\right) ,\ \  
   \beta = \left(\begin{array}{cc}  0  & 1\\ 1 &0\end{array}\right)\ .
\end{eqnarray}
As we shall see, each of these representations leads to an elegant
(but different) rewrite in the 1+1 light cone coordinates for
space-time. RI leads to an equation with real-valued solutions. RII
leads to an equation that corresponds directly to Feynman's
checkerboard model for the 1+1 Dirac equation (Ref. \cite{quantum}).
The lattice paths of Feynman's model are the key to finding
solutions to both versions of the equation. We shall see that these
paths lead to exact solutions to natural discretizations of the
equations.

We now make the translation to light cone coordinates. First
consider RI. Essentially this trick for replacing the complex Dirac
equation by a real equation was suggested to one of us by
V. A. Karmanov \cite{Karmanova}. Using this representation, the Dirac
equation is
\begin{equation}
i\,\frac{\partial \psi}{\partial t} =\left(\pmatrix { i & 0\cr 0 &
-i}\,\frac{\partial}{\partial x} +\pmatrix {0 & -i\cr i & 0}\right)
\psi\end{equation}
whence
\begin{equation}
\frac{\partial \psi}{\partial t} =\left(\pmatrix { 1 & 0\cr 0 &
-1}\,\frac{\partial}{\partial x} +\pmatrix {0 & -1\cr 1 & 0}\right)
\psi\ .
\end{equation}
If $\psi = \pmatrix {\psi_1 \cr \psi_2}$ where $\psi_1$ and $\psi_2$
are real-valued functions of $x$ and $t$, then we have
\begin{equation}
\pmatrix{ -\psi_2 \cr \psi_1} = \pmatrix {\,\frac{\partial \psi_1}
{\partial t}- \frac{\partial \psi_1}{\partial x}\cr \frac{\partial
\psi_2}{\partial t}+ \frac{\partial \psi_2}{\partial x}} \ .
\label{eqAA}
\end{equation}
Now the light cone coordinates of a point $(x,t)$ of space-time are
given by $[r,\ell]=[\frac{1}{2}\,(t+x),\,\frac{1}{2}\,(t-x)]$ and hence the
Dirac equation becomes
\begin{equation}
\pmatrix{ -\psi_2 \cr \psi_1} =
\pmatrix {\frac{\partial \psi_1}{\partial \ell}\cr
\frac{\partial \psi_2}{\partial r}}\ .
\end{equation}
{\bf  Remark.} It is of interest to note that if we were to write
$\psi=\psi_1+i\psi_2$, then the Dirac equation in light cone
coordinates takes the form $D\psi=i\psi$ where
$D(\psi_1+i\psi_2)=\frac{\partial\psi_1}{\partial \ell} +i\,\frac{\partial
\psi_2}{\partial r}$. In any case, we shall refer to Eq. (\ref{eqAA}) as
the RI Dirac Equation

Now, let us apply the same consideration to the second
representation RII. The Dirac equation becomes
\begin{equation}
i\,\frac{\partial \psi}{\partial t} =\left(\pmatrix { i & 0\cr 0 &
-i}\,\frac{\partial}{\partial x} +\pmatrix {0 & 1\cr 1 & 0}\right)
\psi\ .
\end{equation}
Thus
\begin{equation}
\frac{\partial \psi}{\partial t} =\left(\pmatrix { 1 & 0\cr 0 &
-1}\,\frac{\partial}{\partial x} +\pmatrix {0 & -i\cr -i & 0}\right)
\psi \ .
\end{equation}
Hence
\begin{equation}
\pmatrix{ -i\psi_2 \cr -i\psi_1} =
\pmatrix {\frac{\partial \psi_1}{\partial l}\cr
\frac{\partial \psi_2}{\partial r}} \ .
\label{eqBB}
\end{equation}
We shall call (Eq. \ref{eqBB}) the RII Dirac equation.

\section{Discrete Calculus and Solutions to the Dirac Equation}

Suppose that $f=f(x)$ is a function of a variable $x$. Let $\Delta$
be a fixed non-zero constant. The discrete derivative of $f$ with
respect to $\Delta$ is then defined by the equation
\begin{equation}
D_{\Delta}f(x)= \frac{f(x+\Delta)- f(x)}{\Delta} \ .
\end{equation}

Consider the function
\begin{equation}
 x^{(n)}=x(x-\Delta)(x-2\Delta)\cdot\cdot\cdot (x-(n-1)\Delta)\ .
\end{equation}
{\bf Lemma.}
\begin{equation}
 D_{\Delta}x^{(n)}=nx^{(n-1)} \ .
\end{equation}
{\bf Proof.}
\[
 (x +\Delta)^{(n)}-x^{(n)} =\]
\[(x+\Delta)(x)(x-\Delta)\cdot\cdot\cdot(x-(n-2)\Delta)
-(x)(x-\Delta)\cdot\cdot\cdot(x-(n-2)\Delta)(x-(n-1)\Delta) =\]
\[[(x +\Delta)-(x-(n-1)\Delta)]x^{(n-1)}=n\Delta x^{(n-1)}\ .\]
Thus
\begin{equation}
D_{\Delta}x^{(n)}=\frac{n\Delta x^{(n-1)}}{\Delta}=nx^{(n-1)}\ .
\end{equation}
We are indebted to Eddie Grey for reminding us of this fact
\cite{Grey}.

Note that as $\Delta$ approaches zero $x^{(n)}$ approaches $x^n$,
the usual $n^{th}$ power of $x$. Note also that
\begin{equation}
 \frac{x^{(n)}}{\Delta^n n!}= C_n^{x/\Delta}
\end{equation}
where
\begin{equation}
 C_n^{z} = \frac{z(z-1)\cdot\cdot\cdot(z-n+1)}{n!}
\end{equation}
is a (generalized) binomial coefficient. Thus
\begin{equation}
 \frac{x^{(n)}}{n!} = C_n^{x/\Delta} \ .
\end{equation}

With this formalism  in hand, we can express functions whose
combination will yield solutions to discrete versions of the RI and
RII Dirac equations described in the previous section. After
describing these solutions, we shall interpret them as sums over
lattice paths.

To this end, let  $\partial_{\Delta}/\partial r$ and
$\partial_{\Delta}/\partial \ell$ denote discrete partial derivatives
with respect to variables $r$ and $\ell$. Thus
\begin{equation}
\frac{\partial_{\Delta} f}{\partial r} = \frac{f(r+\Delta, \ell)
-f(r,\ell)}{\Delta};\qquad \frac{\partial_{\Delta} f}{\partial \ell} = 
\frac{f(r,\ell+\Delta)-f(r,\ell)}{\Delta}\ .
\end{equation}
Define the following functions of $r$ and $\ell$
\begin{eqnarray}
\psi_R^{\Delta}(r,\ell) &=&
\Sigma_{k=0}^{\infty}
(-1)^k\frac{r^{(k+1)}}{(k+1)!}\ \frac{\ell^{(k)}}{k!} 
\nonumber\\
\psi_L^{\Delta}(r,\ell) &=&
\Sigma_{k=0}^{\infty}(-1)^k 
\ \frac{r^{(k)}}{k!}\ \frac{\ell^{(k+1)}}{(k+1)!} \\
\psi_0^{\Delta}(r,\ell) &=& 
\Sigma_{k=0}^{\infty} (-1)^k\ \frac{r^{(k)}}{k!}
\ \frac{\ell^{(k)}}{k!}\ . \nonumber
\end{eqnarray}
Note that as $\Delta \rightarrow 0$, these functions approach the
limits:
\begin{eqnarray}
\psi_R(r,\ell) &=& \Sigma_{k=0}^{\infty} (-1)^k\ \frac{r^{k+1}}{k!}
\ \frac{\ell^{k}}{k!}\nonumber \\
\psi_L(r,\ell) &=& \Sigma_{k=0}^{\infty} (-1)^k\ \frac{r^{k}}{k!}
\ \frac{\ell^{k+1}}{k!}
\\
\psi_0(r,\ell) &=& \Sigma_{k=0}^{\infty} (-1)^k\ \frac{r^{k}}{k!}
\ \frac{\ell^{k}}{k!} \ .\nonumber
\end{eqnarray}
Note also, that if $r/\Delta$ and $\ell/\Delta$ are {\it positive
integers}, then $ \psi_R^{\Delta}$, $\psi_L^{\Delta}$ and $
\psi_0^{\Delta}$ are {\it finite sums} since $\frac{x^{n}}{n!}=
\Delta^n C_n^{x/\Delta}$ will vanish for sufficiently large $n$ when
$x/\Delta$ is a sufficiently large integer.

Now note the following identities about the derivatives of these
functions
\begin{eqnarray} 
\frac{\partial_{\Delta}\psi_R^{\Delta}}{\partial r} &=& \psi
_0^{\Delta} ,\ \ \ \frac{\partial_{\Delta}\psi_0^{\Delta}}{\partial
r} = -\psi _L^{\Delta}\nonumber \\
\frac{\partial_{\Delta}\psi_L^{\Delta}}{\partial r} &=& \psi
_0^{\Delta} ,\ \ \ \frac{\partial_{\Delta}\psi_0^{\Delta}}{\partial
\ell} = -\psi _R^{\Delta} \ . 
\end{eqnarray} 
With $\Delta = 0$, these can be regarded as continuum derivatives.

We can now produce solutions to both the RI and the RII Dirac
equations. For RI, we shall require
\begin{equation}
 \frac{\partial\psi_1}{\partial \ell} = \psi _2
,\ \ \ \frac{\partial\psi_2}{\partial r} = -\psi _1\ .
\end{equation}
We shall omit writing the $\Delta$'s in those equations, since all
these calculations take the same form independent of the choice of
$\Delta$. Of course for finite $\Delta$ and integral $r/\Delta$,
$\ell/\Delta$ these series produce discrete calculus solutions to the
equations.

Let
\begin{equation}
\psi_1= \psi_0-\psi_L \qquad \ \
\psi_2= \psi_0+\psi_R \ .
\end{equation}
It follows immediately that this gives a solution to the RI Dirac
equation.

Similarly, if we let
\begin{equation}
\psi_1= \psi_0-i\psi_L \qquad \ \ \psi_2= \psi_0-i\psi_R
\end{equation}
then
\begin{equation} 
\frac{\partial\psi_1}{\partial \ell} = -i\psi _2
,\ \ \ \ \ \frac{\partial\psi_2}{\partial r} = -i\psi _1\ .
\end{equation}
This gives a solution to the RII Dirac equation.

In the next section we consider the lattice path interpretations of
these solutions.

\section{Lattice Paths}

In this section we interpret the discrete solutions of the Dirac
equation given in the previous section in terms of counting lattice
paths. As we have remarked in the previous section, the solutions
are built from the functions $\psi_0$, $\psi_R$ and $\psi_L$. These
functions are finite sums when $r/\Delta$ and $\ell/\Delta$ are
positive integers, and we can rewrite them in the form
\begin{eqnarray}
\psi_R(r,\ell) &=& \Sigma_{k=0}^{\infty}
(-1)^k\Delta^{2k+1}C_{k+1}^{r/\Delta}C_k^{\ell/\Delta}\nonumber \\
\psi_L(r,\ell) &=& \Sigma_{k=0}^{\infty}
(-1)^k\Delta^{2k+1}C_{k}^{r/\Delta}C_{k+1}^{\ell/\Delta} \\
\psi_0(r,\ell) &=& \Sigma_{k=0}^{\infty}
(-1)^k\Delta^{2k+1}C_k^{r/\Delta}C_k^{\ell/\Delta}\nonumber
\end{eqnarray}
where
\begin{equation}
C_n^{z} = \frac{z(z-1)\cdot\cdot\cdot(z-n+1)}{n!}
\end{equation}
denotes the choice coefficient.

\vspace{.5cm}
\begin{figure}[htb]
\begin{center} 
\leavevmode
{\epsfxsize=3.75truein \epsfbox{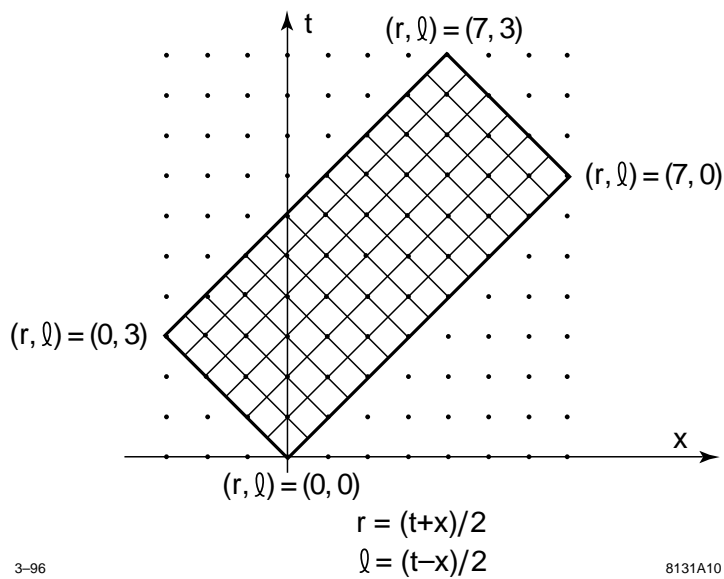}}
\end{center}
\caption{Rectangular lattice in Minkowski space-time.}
\label{fig1}
\end{figure}

We are thinking of $r$ and $\ell$ as the light cone coordinates
$r=\frac{1}{2}\,(t+x)$, $\ell=\frac{1}{2}\,(t-x)$. Hence, in a standard
diagram for Minkowski space-time, a pair of values $[r,\ell]$
determines a rectangle with sides of length $\ell$ and $r$ on the left
and right pointing light cones. (We take the speed of light $c=1$.)
This is shown in Figure. 1.

Clearly, the simplest way to think about this combinatorics is to
take $\Delta =1$. If we wish to think about the usual continuum
limit, then we shall fix values of $r$ and $\ell$ and choose $\Delta$
small but such that $r/\Delta$ and $\ell/\Delta$ are integers. The
combinatorics of an $r\times \ell$ rectangle with integers $r$ and
$\ell$
is no different in principle than the combinatorics of an
$(r/\Delta)\times (\ell/\Delta)$ rectangle with integers $r/\Delta$ and
$\ell/\Delta$. Accordingly, we shall take $\Delta = 1$ for the rest of
this discussion, and then make occasional comments to connect this
with the general case.

\vspace{.5cm}
\begin{figure}[htb]
\begin{center} 
\leavevmode
{\epsfxsize=3.75truein \epsfbox{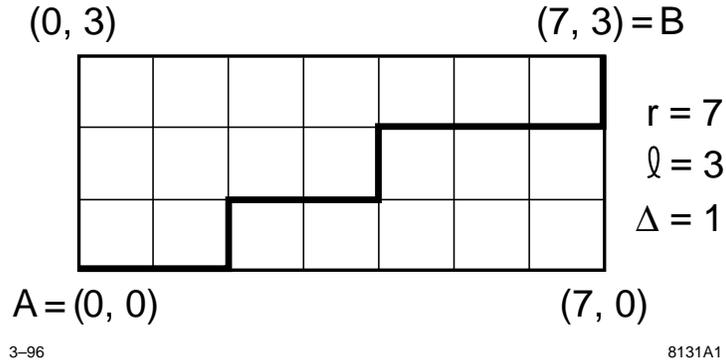}}
\end{center}
\caption{An example of a path in the light-cone
rectangle.}
\label{fig2}
\end{figure}

Finally, for thinking about the combinatorics of the $r\times \ell$
rectangle, it is useful to view it turned by $45^o$ from its
light-cone configuration. This is shown in Fig. 2.
We shall consider lattice paths on the $r\times \ell$
rectangle from $A=[0,0]$ to $B=[r,\ell]$. Each step in such a path
consists in an increment  of either the first or the second light
cone coordinate. The ``particle'' makes a series of ``left or
right'' choices to get from A to B. In counting the lattice paths we
shall represent {\it left} and {\it right} by
\bigskip

\hskip2in\vbox{\epsfxsize=.75truein\epsfbox{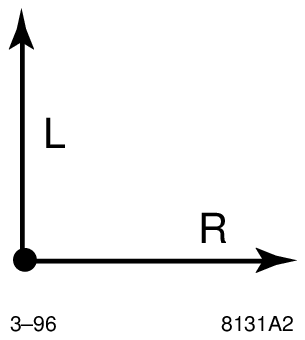}}

\noindent
(Left is vertical in the rotated representation.) 
Now notice that a lattice path has two types of corners:

\hskip1.5in\vbox{\epsfxsize=1.75truein \epsfbox{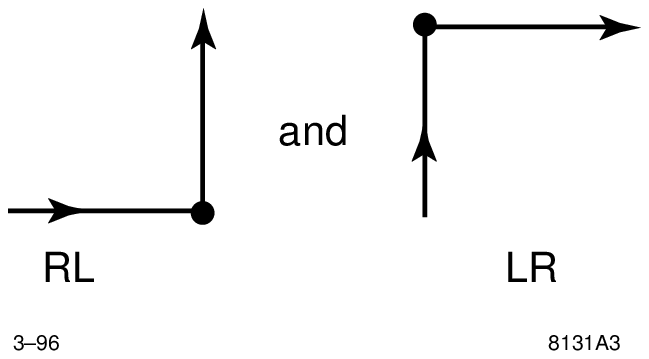}}

We can count RL corners by the point on the L axis where the path
increments. We can count LR corners by the point on the R axis where
the path increments. A lattice path is then determined by a choice
of points from the L and R axes. More specifically, there are paths
that begin in R (go right first) and end in L, begin in L and end in
R, begin in L and end in L, begin in R and end in R. We call these
paths of type RL, LR, LL and RR respectively. (Note that a RL corner
is a two-step path of type RL and that an LR corner is a two step
path of type LR.) It is easy to see that an RL path involves $k$
points from the R axis and $k+1$ points from the L axis, an LR path
involves $k+1$ points from the R axis and k points from the L axis,
while an LL or RR path involves the choice of $k$ points from each
axis. See Figure 3 for examples.

\vspace{.5cm}
\begin{figure}[htb]
\begin{center} 
\leavevmode
{\epsfxsize=3.5truein \epsfbox{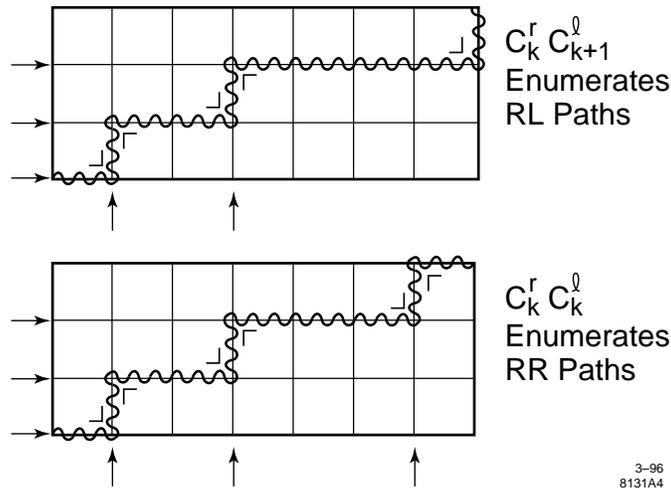}}
\end{center}
\caption{Showing by example that $C_k^rC_{k+1}^\ell$ enumerates RL
paths and $C_k^rC_k^\ell$ enumerates RR paths.}
\label{fig3}
\end{figure}

\noindent
As a consequence, we see that if $\Vert XY\Vert$ denotes the number
of paths from A to B of type XY, then
\begin{eqnarray}
\Vert RL\Vert &=& \Sigma_kC_k^rC_{k+1}^\ell \nonumber \\
\Vert LR\Vert &=& \Sigma_kC_{k+1}^rC_k^\ell \\
\Vert RR\Vert &=& \Vert LL\Vert=\Sigma_kC_k^rC_k^\ell\ . \nonumber
\end{eqnarray}
We see, therefore, that our functions  $\psi_0$, $\psi_R$ and
$\psi_L$ can be regarded as weighted sums over these different types
of lattice path. In fact, we can re-interpret $(-)^k$ in terms of
the number of corners (choices) in the paths:
\newpage

\begin{eqnarray*}
RR &\Rightarrow& 2k   \ \hbox{corners} \\
LR &\Rightarrow& 2k+1 \ \hbox{corners} \\
RL &\Rightarrow& 2k+1 \ \hbox{corners} \\
LL &\Rightarrow& 2k   \ \hbox{corners} \ .
\end{eqnarray*}
Hence if $N_c(XY)$ denotes the number of paths with $c$ corners of
type XY then 
\begin{eqnarray}
\psi_0 &=& \Sigma_c (-1)^{\frac{c}{2}}N_c(LL) = \Sigma_c (-1)
^{\frac{c}{2}}\,N_c(RR)\nonumber \\
\psi_R &=& \Sigma_c (-1)^{\frac{c-1}{2}}\,N_c(LR) \\
\psi_L &=& \Sigma_c (-1)^{\frac{c-1}{2}}\,N_c(RL)\ . \nonumber
\end{eqnarray}
From the point of view of the solution to the RI Dirac equation
($\psi_1=\psi_0-\psi_L$, $\psi_2=\psi_0+\psi_R$) it is an
interesting puzzle in discrete physics to understand the nature of
the negative case counting that is entailed in the solution. (An
attempt has been made by one of us to interpret this in terms of
spin or particle number conservation in the presence of random
electromagnetic fluctuations producing the paths
\cite{NoyesEssays}.) The signs do not appear to come from local
considerations along the path.

The RII Dirac solution gives a different point of view. Here
$\psi_1=\psi_0-i\psi_L$, $\psi_2=\psi_0-i\psi_R$. Taken the hint
given by the appearance of $i$, we note that $i^{2k}=(-)^k$ while
$i^{2k+1} = (-1)^ki$. Thus
\[
\psi_1 = \Sigma_c(-i)^cN_c(R) \qquad
\psi_2 = \Sigma_c(-i)^cN_c(L) \]
where $N_c(R)$ denotes the number of paths that start to the right
and have $c$ crossings, while $N_c(L)$ denotes the number of paths
that start to the left and have $c$ crossings. This shows that our
solution in the RII case is precisely in line with the amplitudes
described by Feynman and Hibbs (Ref. \cite{quantum}) for their
checkerboard model of the Dirac propagator. See also H. A. Gersch
\cite{Gersch} and Ref. \cite{Jacobson} for the relationship of the
Feynman model to the combinatorics of the Ising model in statistical
mechanics.

Returning now to the RI equation, we see that $(-i)^cN_c$ gives the
clue to the combinatorics of the signs. In our RI formulation, no
complex numbers appear and none are needed if we take a
combinatorial interpretation of $i$ as an operator on ordered pairs:
$i[a,b] = [-b,a]$. Then we can think of a ``pre-spinor'' in the form
of a labeled $\frac{\pi}{2}$ angle associated to each corner:

\vspace{.5cm}
\begin{figure}[htb]
\begin{center} 
\leavevmode
{\epsfxsize=2truein \epsfbox{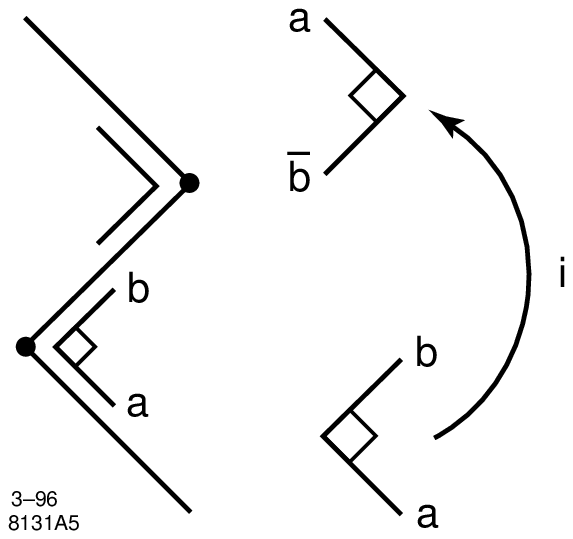}}
\end{center}
\end{figure}

\noindent
As the particle moves from corner to corner its pre-spinor is
operated on by $i$. There is a combination of one sign change and
one change in order. The total sign change from the beginning of the
path to the end documents the positivity or negativity of the count.

\section{Epilogue}

If we had started by saying (in the RI case) we had a simple
solution for the Dirac equation (discretized) using nothing but {\it
bit-strings} (L,R choice sequences) and appropriate signs, then it
would have been natural to ask: How are these signs justified on the
basis of a philosophy of bit-strings? In retrospect we can answer:
This pattern of signs is very simple, but not (yet) to be deduced
from the notion of a distinction alone. Nevertheless, it does arise
naturally from the simple structures that are available at that
primitive level. The $i$ operator ($i[a,b] = [-b,a]$) does not
involve anything more sophisticated that the idea of exchanging the
labels on the two sides of a distinction followed by the flipping of
a label on a given side:
\newpage

\vspace{.5cm}
\begin{figure}[htb]
\begin{center} 
\leavevmode
{\epsfxsize=3truein \epsfbox{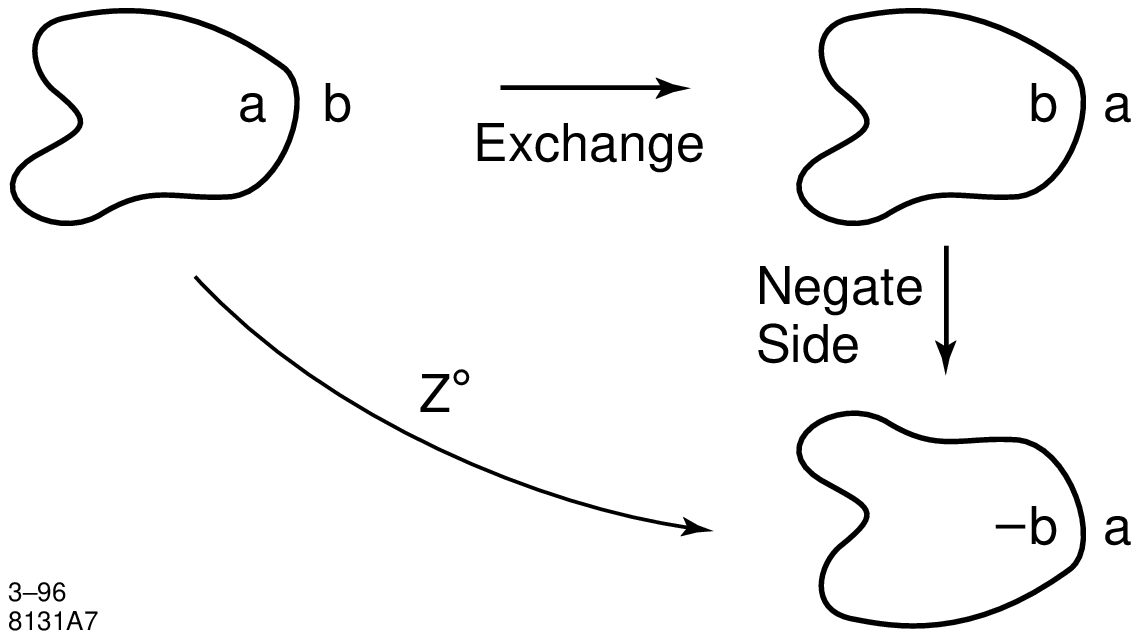}}
\end{center}
\end{figure}

\noindent
as is discussed elsewhere \cite{Kauffman,SpecialRelativity}.  A
choice sequence such as
\bigskip

\hskip1.5in\vbox{\epsfxsize=2.75truein\epsfbox{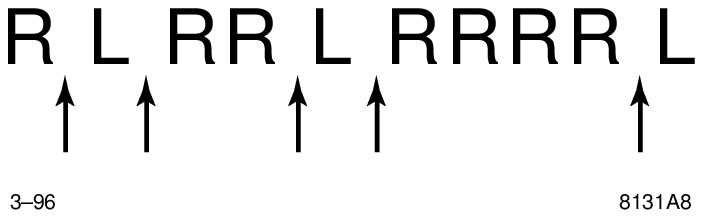}}

\noindent
has ``corners'' wherever R meets L or L meets R. We have
characterized these corners into two types RL and LR:
\bigskip

\hskip1.5in\vbox{\epsfxsize=2.75truein\epsfbox{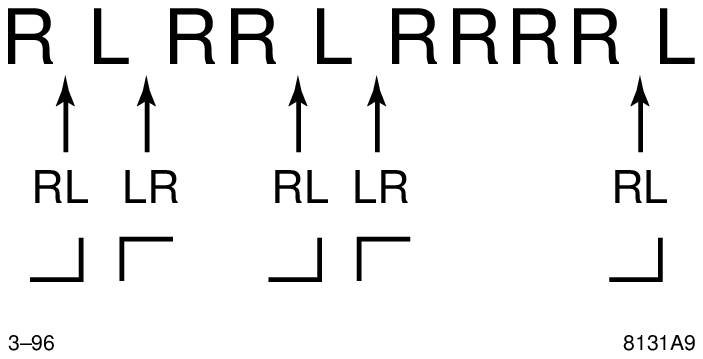}}

\noindent
We then enumerate the choice sequences in terms of lattice paths in
Minkowski space and the solutions to the Dirac equation emerge,
along with a precursor to spin and the role of $i=\sqrt{-1}$ in
quantum mechanics. We have shown exactly how this point of view
interfaces with Feynman's Checkerboard.

Corners in the bit-string sequence alternate from RL to LR and from
LR to RL. The moral of Feynman's $(-i)^c$ where $c$ is the number of
corners is that this alteration should be regarded as an elementary
rotation:
\newpage

\vspace{.5cm}
\begin{figure}[htb]
\begin{center} 
\leavevmode
{\epsfxsize=3.75truein \epsfbox{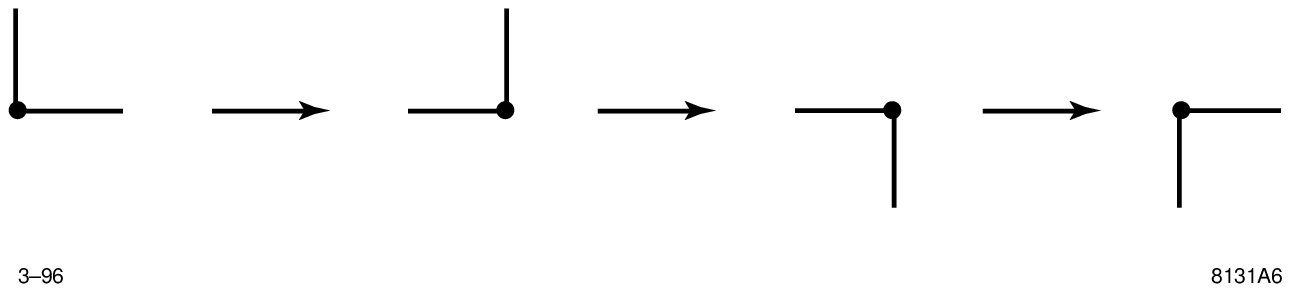}}
\end{center}
\end{figure}

\noindent
One may wonder, why does this simple combinatorics occur in a level
so close to the making of one distinction, and yet implicate fully
the solutions to the Dirac equation in continuum 1+1 physics?! We
cannot begin to answer such a question except with another question:
If you believe that simple combinatorial principles underlie not
only physics and physical law, but the generation of space-time
herself, then these principles remain to be discovered. What are
they? What are these principles? It is no surprise to the
mathematician that $i$ ends up as central to the quest. For $i$ is a
strange amphibian not only neither 1 nor $-1$, $i$ is neither discrete
nor continuous, not algebra, not geometry, but a communicator of
both. In this essay we have seen the beginning of a true connection
of discrete and continuum physics.

The continuum version of our theory merges the paths on the lattice
to a sum over all possible paths on an infinitely divided rectangle
in Minkowski space-time. The individual paths disappear into the
values of the series $ \psi_0 = \Sigma_{k=0}^{\infty} (-1)^k
\frac{r^{k}}{k!}\,\frac{\ell^{k}}{k!} $, $
\psi_L=\Sigma_{k=0}^{\infty}(-1)^k
\frac{r^{k}}{k!}\,\frac{\ell^{k+1}}{(k+1)!}$ , $
\psi_R=\Sigma_{k=0}^{\infty}(-1)^k
\frac{r^{k+1}}{(k+1)!}\,\frac{\ell^{k}}{k!} $. Here we have a glimpse
of the possibilities inherent in a complete story of discrete
physics {\it and} its continuum limit. The continuum limit will be
seen as a {\it summary} of the real physics. It is a way to view,
through the glass darkly, the crystalline reality of simple quantum
choice.
\newpage

\section*{Acknowledgments}
As is discussed more fully in Ref. \cite{NoyesEssays}, this line of
investigation started thanks to correspondence between V. A.
Karmanov and I. Stein about the possibility of relating the
Feynman-Hibbs suggestion to the Stein model,
\cite{SteinSeminars,SteinPapers,SteinEssays,SteinConcept} and a
comment by D. O. McGoveran that an approximation suggested by
Karmanov was already the exact result. Unfortunately these three
authors could not come to consensus with each other and/or HPN as to
how to present the work. Several drafts were also criticized by C.
W. Kilmister and J. C. van den Berg. The work presented here follows
a somewhat different approach, but has drawn heavily on the
experience gained in collaboration and discussion with all five of
these scientists.

\end{document}